\documentstyle[preprint,aps]{revtex}
\begin{document}
\title{Stable and Metastable vortex states and the first order
transition across the peak effect region in weakly pinned 2H-NbSe$_2$}
\author{G. Ravikumar$^*$ and  V.  C. Sahni}
\address{Technical Physics and Prototype Engineering Division,
Bhabha Atomic Research Centre, Mumbai-400085, India}
\author{A. K. Grover and S. Ramakrishnan}
\address{Dept. of Condensed Matter Physics and Materials Science, Tata
Institute of Fundamental Research, Mumbai-400005, India}
\author{P. L. Gammel, D. J. Bishop and E. B\"ucher}
\address{Bell  Laboratories,  Lucent Technologies, Murray Hill, New
Jersey 07974}
\author{M. J. Higgins}
\address{NEC  Research Institute, 4 Independence Way, Princeton, New
Jersey 08540}
\author{S. Bhattacharya$^*$}
\address{NEC Research Institute, 4 Independence Way,  Princeton,  New
Jersey 08540 and Dept. of Condensed Matter Physics and Materials Science, Tata
Institute of Fundamental Research, Mumbai-400005, India}
\maketitle
\begin{abstract}
The  peak  effect in weakly pinned superconductors is accompanied
by metastable vortex states. Each metastable vortex configuration
is characterized by a different critical current  density  $J_c$,
which  mainly  depends  on the past thermomagnetic history of the
superconductor. A recent model [G. Ravikumar {\it et  al},  Phys.
Rev.  B  {\bf  61}, R6479 (2000)] proposed to explain the history
dependent $J_c$ postulates a stable state of vortex lattice with
a critical current density $J_c^{st}$, determined uniquely by  the
field and temperature. In this paper, we present evidence for the
existence  of  the stable state of the vortex lattice in the peak
effect region of $2H-NbSe_2$. It is shown that this stable  state
can  be  reached  from any metastable vortex state by cycling the
applied field by a small amplitude. The minor magnetization loops
obtained by repeated field cycling  allow  us  to  determine  the
pinning  and  "equilibrium" properties of the stable state of the
vortex lattice at a given field  and  temperature  unambiguously.
The  data  imply  the occurence of a first order phase transition
from an ordered phase to a disordered  vortex  phase  across  the
peak effect.
\end{abstract}
\newpage
\noindent
\normalsize
\section{INTRODUCTION}
In  the  presence  of strong pinning, the vortex state of type II
superconductors is usually characterized by the critical  current
density  $J_c(H,T)$ which decreases monotonically with increasing
field $H$ or temperature $T$. In the weakly  pinned  superconductors,
on  the other hand, the interplay between the intervortex interaction
and the flux pinning  produces  an  anomalous  peak  in  $J_c$,  as  a
function  of  both field and temperature\cite{higgins} just below
the normal state boundary (usually designated as the peak  effect
or  PE).  Within the collective pinning description\cite{larkin},
this signifies that the vortex phase undergoes a transition/crossover
from      an      ordered      state      to     a     disordered
state\cite{higgins,satya1,gammel,chandra}. The detailed nature of
this transition,  e.g.,  whether  it  is  a  thermodynamic  phase
transition or not, remains a subject of considerable debate.

One  of  the  key  issues  is  the detection of an anomaly in the
thermodynamic  quantities,  such  as,  specific heat or equilibrium
magnetization  $M_{eq}$. $J_c$ and $M_{eq}$ can be estimated from
the measured irreversible magnetization data of a superconducting
sample \cite{bean,fietz,pc} using the relations,

$$ J_c(H)   = [ M(H\downarrow)  -   M(H\uparrow)   ]/2  g  \mu_0  R,
\eqno(1a)$$
$$ M_{eq}(H) = [ M(H\uparrow)  +  M(H\downarrow)  ]/2,  \eqno(1b)$$

\noindent
where  $M(H\uparrow)$  and $M(H\downarrow)$ are the magnetization
in the increasing (forward) and decreasing (reverse) field cycles
respectively,  $\mu_0  =  4 \pi \times 10^{-7}$ W/A.m, $R$ is the
sample dimension transverse to the applied field  and  $g$  is  a
factor which depends on the sample geometry. Eq. 1 implicitly assumes
that $J_c$ is history independent and is thus uniquely determined
by  the  local  induction  $B$.  However,  across the peak effect
region, the above equations are not valid due to a strong history
dependence  in  $J_c$ 
\cite{satya1,steingart,wordenweber,henderson,fcrev,banerjee,supercool,shampa,roy,degroot}.
Recently,  considerable  efforts  have  gone  into ascertaining the
equilibrium magnetization across the peak effect region, where  an
order-disorder  transition  occurs in the vortex matter. However,
these efforts have met with ambiguous  and  somewhat  conflicting
results.   For  example,  the  construction  of  the  equilibrium
magnetization from the hysteresis loop  by  using  two  different
kinds  of  minor magnetization curves\cite{roy,tenya,stepchange},
results in apparently  different  conclusions.  In  one  case,  a
jump\cite{roy,stepchange} in $M_{eq}$ could be found at the onset
of  the PE,  while the other case shows no increase at all\cite{tenya}.
These differences apparently originate from the  difficulties  in
establishing  an unambiguous and reproducible vortex state due to
a strongly history dependent configuration of the vortex matter in 
the PE region. The different procedures  proposed  to obtain $M_{eq}$  
shall  be discussed in section II.

In  Sec.  III,  we  briefly  discuss  a  recent  phenomenological
model\cite{model},  which  addresses  the  issue  of  the  history
dependent $J_c$ and the metastability in the vortex state through
an  extension  of  the Bean's critical state model\cite{bean}. In
Sec.  IV, we present an experimental method based on the ideas of
the model\cite{model} to obtain a unique "stable" vortex state in
the PE region, which is independent of the past magnetic  history.
We  propose  that  this  state,  in  effect,  is  the ``stable'' or
``equilibrium''state and evaluate the critical current density and
$M_{eq}$ of this state. We  further  demonstrate  that,  a  sharp
change  in  the equilibrium magnetization (albeit smeared) occurs
across the PE region. These  results  imply  that  an  underlying
first  order phase transition, presumably driven by a competition
between elastic and pinning energies in a situation where thermal
fluctuations are weak, marks the peak  effect. 

\section{MINOR  CURVES AND THE EQUILIBRIUM MAGNETIZATION ACROSS 
THE PEAK EFFECT}
In  the  peak  effect region, the critical current density in the
increasing  field  cycle  $J_c(H\uparrow)$  is  less  than   that
($J_c(H\downarrow)$)    in    the    decreasing    field    cycle
\cite{steingart,wordenweber,supercool} for $H < H_p$, where $H_p$
is the field where $J_c$ is  maximum.  However,  well  below  the
onset  of  the  PE  and at $H > H_p$, $J_c$ is independent of the
magnetic  history.  One  of  its  consequences  is  the  peculiar
behavior  of  the  minor  magnetization  curves, which can not be
reconciled  within  the  critical  state  model\cite{bean}.   For
instance,  a typical minor magnetization curve (type I) initiated
from a field $H < H_p$ in the  PE  region  {\it  saturates}  {\it
without       meeting}       the       reverse      magnetization
curve\cite{supercool,roy,degroot}, as shown in Fig. 1(a).  On  the
other hand, the minor curves (type II) measured by increasing the
field  from  different  points on the reverse magnetization curve
{\it overshoot} the forward curve\cite{supercool,tenya} as  shown
in Fig.1(b). The two types of  anomalous behavior may be contrasted
with the conventional behavior for the minor curves starting
at (a) $H$ $>$ $H_p$ and (b) H$<<$H$_{pl}$, i.e.,for fields well 
below the PE region. The latter catagories of minor curves meet
the  magnetization  envelope,  constituted  by  the  forward  and
reverse curves, as expected from the Bean's critical state model.

A  new  procedure  was proposed by Roy and Chaddah \cite{roy} to 
obtain $M_{eq}$ from the minor magnetization curves of the type I 
by the relation,

$$
M_{eq}(H) = [ M(H+\delta, \uparrow) + M_{ML}(H-\delta, \downarrow) ]/2,
\eqno(2)
$$

\noindent
where  $M(H+\delta, \uparrow)$ is the magnetization at a field $H
+ \delta$ (denoted by point A in Fig. 1(a)) from where the minor curve
is initiated on the forward curve.  $M_{ML}(H-\delta,\downarrow)$
is  the  magnetization on the minor curve at a field $H - \delta$,
where it saturates as indicated by the point B  in  Fig.  1(a).  This
procedure  is  based  on  the implicit assumption that the vortex
state formed on the forward curve is an "equilibrium" state. This
assumption  is  however  inconsistent   with   the   experimental
observation  by Wordenweber, Kes and Tsuei\cite{wordenweber}, who
showed that both current  cycling  and  field  cycling  processes
eventually establish a vortex state with a $J_c$ higher than that
on  the  forward  curve.  Such  an observation indicates that the
vortex state formed on the forward curve is metastable in nature.

Tenya  {\it  et  al}\cite{tenya}  have preferred a procedure given
below, which is very similar to the one described above but  using
the minor curves of the type II described in Fig. 1(b):

$$
M_{eq}(H) = [ M(H-\delta, \downarrow) + M_{ML}(H + \delta, \uparrow) ]/2,
\eqno(3)
$$

\noindent
where  $H - \delta$ (point C in Fig. 1(b)) is the field from where the
minor curve is initiated on the reverse curve and  $H  +  \delta$
(point D in Fig. 1(b)) is the field where it saturates. $M_{ML}(H
+  \delta, \uparrow)$ is the saturated magnetization value on the
minor curve. This procedure too has the  shortcoming  similar  to
that   in   Eq.   2,  viz.,  the  vortex  state  on  the  reverse
magnetization     curve     is     actually     a      metastable
state\cite{wordenweber,supercool,roy}.  Moreover,  not  only  are
these recipes deficient, they also yield  different  conclusions,
viz.,  an enhancement in equilibrium magnetization is observed in
one case, whereas it is absent in  the  other.  These  ambiguities
point  to  the  need  to  evolve a more satisfactory procedure to
arrive at a unique and  stable  vortex  state  unambiguously  and
determine the equilibrium magnetization assuming the stable state
to be the equilibrium state.
\section{MODEL  FOR  HISTORY  EFFECTS AND METASTABILITY}
Ravikumar   {\it  et  al}\cite{model}  incorporated  the  history
dependence  in  the macroscopic critical current density $J_c$ by
postulating,

$$
J_c(B+\Delta B) = J_c(B) +  (|\Delta  B|/B_r) (J^{st}_c - J_c).
\eqno{(4)}
$$

\noindent
where  the  critical  current  density  $J_c(B)$ is a macroscopic
representation of a particular metastable  configuration  of  the
vortex  lattice  at  a  field $B$. Eq. 4 describes how the vortex
state evolves from one metastable configuration  to  another.  An
important  assumption  of this model is the existence of a stable
vortex state with a critical current density $J_c^{st}$, which is
unique for a given field and temperature. $B_r$ is a  macroscopic
measure  of  metastability and describes how strongly $J_c$ could
be history dependent. In the limit  of  $B_r$  tending  to  zero,
however,  this model reduces to the standard critical state model
for which $J_c$ ($= J_c^{st}$) is  independent  of  the  magnetic
history. It can be seen from Eq. 4 that a metastable vortex state
with $J_c$ $\ne$ $J_c^{st}$, can be driven into a stable state by
merely  oscillating the field by a small amplitude (see Fig. 1 of
the Ref. 20). In the PE region, the energy barriers between different
metastable  vortex  configurations  are  much  greater  than   the
available  thermal  energy. The field cycling allows the vortices
to move and explore the energy landscape and thereby rearrange in
to a vortex configuration closer to the stable state. In the next
section, we will demonstrate this experimentally  and  show  that
the  stable  state obtained is indeed independent of the magnetic
history.

In  the limit $\Delta B \rightarrow 0$, Eq. 4 can be rewritten in
the form,

$$
\pm dJ_c/dB = (J^{st}_c - J_c)/B_r,
\eqno{(5)}
$$

\noindent
where  upper  and  lower  signs  are  applicable  in the cases of
increasing and decreasing local field $B$, respectively. In  each
case, the  $J_c(B)$  can  be  obtained  by solving Eq. 5, provided the
functional form of  $J_c^{st}(B)$  and  $B_r(B)$  are  known.  We
assume  for $J_c^{st}(B)$ and $B_r(B)$ the following forms used
in Ref. \cite{model}  for  calculating  the  minor  magnetization
curves:

$$
J^{st}_c (B)  =  J_{c1}(1  - B/\mu_0 H_1) + J_{c2} e^{-(B -
\mu_0 H_p)^2/2\mu_0 H_W^2}
\eqno{(6)}
$$
and
$$
B_r(B) \approx (B - \mu_0 H_{low})^m (\mu_0 H_p - B)^n~
for~H_{low}<B/\mu_0<H_p $$
$$
\approx 0~~~~~~~~~~~~~~~~~~~~~~~~~~~~~~~~~~~ otherwise
\eqno{(7)},
$$

\noindent
The  first  term  in  Eq. 6 is the field dependence of $J_c^{st}$
well below the peak and the second  term  reflects  the  peak  in
$J_c^{st}$  $vs$ $B$. $B_r(B)$ in Eq. 7 accounts for the observed
history dependence in $J_c$ in the PE region. $B_r =  0$  in  the
field  ranges $H < H_{low}$ and $H > H_p$ signifies that $J_c$ is
independent  of  the magnetic  history  and  is   always   equal   to
the $J_c^{st}$.  For  the  two  limiting cases, $H < H_{low}$ and $H >
H_p$, the intervortex  interaction and the flux pinning  are  dominant
respectively  and  therefore the stable state is readily accessed
by the vortex lattice. The values of the different parameters used in
this paper are listed in the caption of Fig. 2. $J_c(H \uparrow)$
[$J_c(H \downarrow)$] is calculated by numerically solving Eq.  5
with  the upper (lower) sign  with  the  initial  condition  $J_c(H
\uparrow)$ [$J_c(H \downarrow)$] $=$ $J_c^{st}(H)$ at some  field
below  $H_{low}$  (above  $H_p$).  In  Fig.  2(a),  we present an
evaluation of $J_c(H \uparrow)$  and  $J_c(H  \downarrow)$  which
obey   the   inequality   $J_c(H   \uparrow)   <   J_c^{st}(B)  <
J_c(H\downarrow)$. It was earlier  interpreted  that  the  vortex
state formed on the decreasing field cycle is a {\it supercooled}
disordered  state\cite{supercool}.  In  other  words,  the vortex
state formed in decreasing field (from above $H_p$)  retains  the
memory  of  the  vortex correlations from the previous fields. In
analogy, we can  argue  that  the  vortex  state  formed  on  the
increasing field cycle is a {\it superheated} ordered state. Both
of  these  states are metastable in nature. As argued above, they
can be driven into a stable state  by  oscillating  the  external
field by a small amplitude.

The   magnetization   hysteresis  loop  corresponding  to  $J_c(H
\uparrow)$ and $J_c(H \downarrow)$ are shown in Fig.  2(b).  Note
the asymmetry in the hysteresis, usually observed in experiments.
For  a comparison, we also plot the magnetization hysteresis loop
one would obtain within the framework of  Bean's  critical  state
model  with  $J_c  =  J_c^{st}$  (applicable  in  the  limit $B_r
\rightarrow 0$) which is symmetric in the forward and  reverse  field
cycles,  as  shown by the dotted line in Fig. 2(b). Details of the
magnetization calculation are described in  Ref.  20.  The  minor
magnetization curves of the types I and II calculated in the slab
geometry, are shown in Fig. 3(a) and Fig. 3(b) respectively. They
clearly  mimic  the  behavior  seen  in  experiments.  We assumed
$M_{eq}(H) = 0$, in calculating these  magnetization  curves.  We
note  that  the  calculated  curves  in Fig. 2 and Fig. 3 are not
quantitative  fits  to  experimental  data, they only  serve   to
illustrate the qualitative features of the observed data.

In   Fig.   3(c),  we  show  $M_{eq}^*(H)$  determined  from  the
calculated minor curves of the type I and type II following Eq. 2
and Eq. 3, respectively. The test of the self-consistency of these
procedures lies in reproducing the original form ($M_{eq}  =  0$)
assumed in the calculation. $M_{eq}^*(H)$ obtained from these two
procedures  are not only inconsistent with each other, but, also, do
not  conform  to  the  original   assumption   that   $M_{eq}   =
0$\cite{note}. The procedure of Eq. 2 indeed produces a peak like
structure  in  $M_{eq}^*(H)$ which has been shown earlier from an
analysis of experimental data in $2H-NbSe_2$ following  the  same
recipe\cite{stepchange}.  On  the  other  hand,  the use of Eq. 3
proposed by Tenya {\it et al}\cite{tenya} yields no variation  in
$M_{eq}^*$ vs $H$ across the PE region. Fig. 3(c) illustrates the
unreliable  and ambiguous nature of these recipes noted above and
thus points to the need for a consistent  approach  in  order  to
overcome their difficulties.

\section{EXPERIMENTAL RESULTS AND DISCUSSION}
In  this section, we will show experimentally that repeated field
cycling    drives    any   metastable   state   into   a   stable
state, which is unique at a given field\cite{model}. We study  the
minor hysteresis loops traced by repeated field cycling and infer
from  these  measurements the critical current density $J_c^{st}$
and the equilibrium magnetization $M_{eq}$ of the stable state.

{\it DC} magnetization measurements have been carried out using a
Quantum  Design (QD) Inc. SQUID magnetometer (Model MPMS5) in the
peak effect region  of  a  $2H$-$NbSe_2$  single  crystal  ($T_c$
$\approx$  7.25  K)  with  the field applied parallel to its {\it
c}-axis. The crystal is of approximate dimensions  ($a  \times  b
\times  c$) $4 mm \times 5 mm \times 0.43 mm$. As stated earlier,
the  peak  effect  in  $J_c$  is  manifested  as  the   anomalous
enhancement  in  the  magnetization hysteresis (c.f. Fig. 1). The
magnetization  hysteresis   has   been   studied   at   different
temperatures  from 6.7 to 6.95K. Magnetization hysteresis data at
6.95K was measured using a 2 cm full scan length, and the data at
the  other  temperatures  was  obtained   using   the   half-scan
technique\cite{fcrev,halfscan}  to avoid artefacts arising due to
field inhomogeneity experienced by  the  sample  along  the  scan
length.  In the temperature range investigated, $J_c$ at the peak
field $H_p$ decreases with decreasing temperature (see Table 1).

Fig.  4(a)  depicts  a  part  of  the  hysteresis  loop at 6.95K,
constituting $M$ $vs$ $H$ curves in the increasing (forward)  and
decreasing  (reverse)  field  cycles  measured with a 30 sec wait
time at each field. We identify the onset field $H_{pl}^+$ of the
PE on the forward curve, where $M$ begins to decrease sharply. The
field $H_p$ marks the field at which magnetization hysteresis  is
maximum.  In  Fig.  4(a),  we  show the points A, B, C and D from
where the minor hysteresis loops are initiated. A(C) and B(D) are
at a field $H < H_{pl}^+$ ($H > H_{pl}^+$)  on  the  forward  and
reverse curves, respectively. Minor hysteresis loops starting from
both  forward and reverse curves are recorded at different fields
(spanning the peak region) by repeatedly cycling the field  by  a
small  amplitude  $\Delta  H$.  The interval $\Delta H$ is chosen
such that it is above the threshold field required to reverse the
direction of the shielding currents throughout the  sample.  From
the critical state model, we understand that magnetization values
must  always  remain  confined  within  the  forward  and reverse
magnetization   curves,   which   constitute   the   so    called
magnetization  envelope.  Further,  the  $M-H$ loop in each field
cycle must retrace itself.

In  Fig. 4(b), we show the minor hysteresis loops (MHLs) measured
by repeatedly cycling the  field,  starting  at  point  A  ($H  <
H_{pl}^+$)  on  the  forward curve. These MHLs in different field
cycles retrace each other indicating  that  the  $J_c$  does  not
change with field cycling. Therefore, we conclude that the vortex
state  is  in a stable configuration. In contrast, the MHLs shown
in Fig. 4(c) (continuous line with data points omitted)  starting
at  B  ($H  <  H_{pl}^+$) on the reverse curve, show shrinkage 
effects, with each successive field cycle and finally MHL collapses
into the minor  loop
started  from  point  A (open circles) which is replotted in Fig.
4(c). This suggests that the vortex configuration at point  B  is
metastable with a $J_c > J_c^{st}$. Repeated field cycling causes
the $J_c$ to fall towards the stable stable value as reflected in
the reduction of the  width  of  the  MHL  with each successive field
cycle. It is remarkable that the minor loops starting from both A
and B merge into precisely  the  same  loop  within the experimental
accuracy.  This  clearly  reaffirms  the  basic assumption of the
model that there exists  a  unique  stable  state  with  critical
current  density  $J_c^{st}$,  independent  of the initial vortex
state from which it evolves.

We  now focus on the behavior of MHLs which start from a field $H
> H_{pl}^+$. As shown in Fig. 4(d), the  behavior  of  the  minor
loops starting at point C is quite different from those started at
point  A.  The increasing field leg of the MHL moves away from the
forward  magnetization  curve in the first field cycle itself and
remains outside the magnetization envelope for  subsequent  field
cycles.  This  clearly  suggests  that,  for  $H > H_{pl}^+$, the
vortex configuration even on the forward magnetization  curve  is
metastable. However, the behavior of the MHLs starting at point D
on  the  reverse  magnetization  curve is very similar to the 
behavior of those
that start  at  point  B,  i,e.,  the  MHL  shrinks  with  each
successive  field  cycle (continuous line in Fig. 4(e)). The data
in Fig. 4(d) is replotted in Fig. 4(e) (open circles connected by
dotted line), which suggests that the MHLs starting from  both  C
and  D  collapse  into  the  same  final loop (MHL). Firstly, the
data in Fig. 4 clearly suggest the metastable nature of the vortex
configuration for fields above $H_{pl}^+$ both on the forward and
the reverse  magnetization  curves.  Further, the eventual MHL
obtained on repeated field cycling is independent of the initial vortex
configuration. We note that the metastable state on  the  forward
magnetization  curve  settles  into  the stable state much faster
than that on the reverse curve. This might imply that the  vortex
configuration  on  the  increasing  field  cycle is closer to the
equilibrium configuration.

The   data  in  Fig.  4  yield  the  following inequalities for the
critical currents in the different field ranges: (i) For $H < H_{pl}^+$,
the vortex  configuration  is  stable  in  the increasing field cycle
while at the same field value, it is  highly  metastable  in  the
decreasing  field cycle. This can be summarized by the inequality
$J_c(H\uparrow) = J_c^{st}(H)  <  J_c  (H\downarrow)$;  (ii)  For
$H_{pl}^+ < H < H_p$, the vortex configurations in both increasing
and decreasing field cycles are  metastable,  with  the  critical
currents  obeying  the inequality, $J_c(H\uparrow) < J_c^{st}(H) <
J_c (H\downarrow)$; (iii) For $H > H_p$, $J_c(H\uparrow) = J_c
(H\downarrow)  =  J_c^{st}(H)$.   These   observations   are   in
accordance  with  the model\cite{model} (cf. Fig. 2(a)). We thus
assert that the Eq. 2,  proposed  by  Roy  and  Chaddah\cite{roy} is
applicable  only  for  $H < H_{pl}^+$. It is unsatisfactory for 
$H_{pl}^+ < H < H_p$,  as  the  vortex  lattice  on the forward 
curve is in a {\it superheated} vortex configuration  which  is  
more  ordered  (but metastable)  than  the stable configuration. 
Eq. 3, as proposed by Tenya {\it et al}\cite{tenya} is not appropriate 
in  any  of  the field  ranges  because  the  vortex states produced 
on the reverse curve are {\it supercooled} vortex
configurations\cite{supercool,royreview}  which are more disordered
than the corresponding stable states.

Fig.  5  shows  the  M-H  loop  at  6.9K constituting the forward and
reverse magnetization curves (dark line with data points omitted)
indicating $H_{pl}^+$ and $H_p$. Note the asymmetry (also seen at
6.95K)  in  the forward and reverse magnetization curves which is
the hall mark of the peak effect. We also measured  the  MHLs  by
repeatedly  cycling the field starting at different points on the
forward and reverse curves. The saturated MHLs are again found to
be independent of the initial vortex state just as for 6.95K. The
locus of magnetization values on the  increasing  and  decreasing
field legs of the saturated MHLs measured at different fields are
also  plotted  in Fig. 5 (open circles connected by dotted line).
This observed behavior is in excellent  qualitative  agreement  
with  that expected  from the model in Ref. 20 (see Fig. 2(b)). 
The locus of saturated magnetization values  corresponds  to  the  
``table"  or the ``quilibrium" vortex configuration at different fields.

Having  established the existence of a history independent stable
state we determine the critical current  density  $J_c^{st}$  and
the equilibrium magnetization $M_{eq}$ state  at  each field from 
the saturated MHL\cite{jc}. $J_c^{st}$ and $M_{eq}$ are given by,

$$ J_c^{st}(H)   = [ M_{st}(H\downarrow)  -   M_{st}(H\uparrow)   ]/2  g  \mu_0  R,
\eqno(8a)$$
$$ M_{eq}(H) = [ M_{st}(H\uparrow)  +  M_{st}(H\downarrow)  ]/2,  \eqno(8b)$$

\noindent
where   $M_{st}(H\uparrow)$  and  $M_{st}(H\downarrow)$  are  the
magnetization values on the increasing and decreasing field  legs
of  the saturated MHL. $J_c^{st}$ vs $H$ and $M_{eq}$ vs $H$ data
at  6.95K  are  plotted  in Fig. 6(a) and Fig. 6(b), respectively.
$M_{eq}$ exhibits a sharp increase between $H_{pl}^+$  and  $H_p$
signifying  an  increase in the equilibrium flux density. This is
reminiscent of the characteristic of $M_{eq}$ across the FLL melting
transition  observed  in  cuprate
superconductors\cite{pastoriza,zeldov}.  We argue that the change
in $M_{eq}$ indicates a first order transition in the FLL from an
ordered  solid  to  a  pinned  amorphous   state\cite{stepchange}
presumably  analogous  to  a  Bragg  Glass to Vortex Glass/pinned
liquid phase transition\cite{giamarchi}. The increase in $M_{eq}$
coincides with the increase in $J_c^{st}$ near the onset  of  the
peak  effect  and  spans  the  field range between $H_{pl}^+$ and
$H_p$. In Fig. 7(a) and 7(b), we present the $M_{eq}$ vs $H$  and
$J_c^{st}$  vs  $H$  data respectively, at 6.9K. We note that the
sharp change in $M_{eq}$ correlates  with  a  sharp  increase  in
$J_c^{st}$  between  $H_{pl}^+$  and  $H_p$.  We also present the
$\Delta M_{eq}$ values  obtained  at  different  temperatures  in
Table  1. 

It  is  important to understand the nature of the vortex state in
the transition region $H_{pl}^+ < H < H_p$. One of  the well known
pictures is  the collective pinning scenario\cite{larkin}, where 
the loss of long range order is expected to permeate uniformly  
throughout the sample. On  the  other hand  
Paltiel {\it et al}\cite{paltiel} have recently proposed a picture where
the disordered phase enters  through  surface  imperfections  and
coexists  near the surface with the ordered phase of the bulk. They
argue that the boundary between the  disordered  region  and  the
ordered  region  moves  into  the  sample  as the temperature (or
field) is increased towards $T_p$ (or $H_p$). Further possibility
is the coexistence  of  ordered  and  disordered  phases,  with  an
intricate  geometrical connectivity of these phases. Irrespective
of the particular picture used,  our  experiments  demonstrate  a
specific  and an unambiguous procedure, viz., subjecting the sample
to a field cycling, to  produce  a  unique  stable  state  (in  a
macroscopic sense) across the peak effect region.

We  consider this stable state as a pinned equilibrium state, and
estimate equilibrium magnetization and the free energy difference
or entropy change when the vortex lattice changes from an ordered
to  an   amorphous   state.   As   per   the   Clausius-Clapeyron
relation\cite{zeldov,welp},  the  entropy  change  per vortex per
inter-layer  distance  $d$  ($\simeq$   4   Angstroms)   in   the
$2H-NbSe_2$ system\cite{stepchange},

$$
\Delta s = -(\Delta M_{eq}/H_p) (dH_{pl}^+/dT) (\phi_0 d/k_B),
$$
\noindent
where $dH_{pl}^+/dT$ $\simeq$ $dH_p/dT$ $\simeq$ $-0.65$ T/K. The
value  of  $\Delta  s$  estimated  at  different  temperatures is
tabulated in Table 1. Incidentally these values are comparable to
the entropy change reported across the FLL melting transition  in
high $T_c$ cuprates.

An  important  question  that  can  arise  is whether the entropy
change can be observed in thermal measurements such  as  specific
heat  vs  temperature.  We  recall  that the metastability in the
vortex state is much greater in  temperature  scans  in  a  fixed
magnetic  field\cite{supercool}.  Repeated  cycling of the field by a
small amplitude may be  necessary  to  produce  the  "stable"  or
"equilibrium"  state  before a thermal measurement is carried out
at each temperature.

\section{CONCLUSIONS}
In  this paper, we have presented a study of the different metastable
vortex configurations occuring in the peak  effect  region  of  a
weakly  pinned  superconductor  $2H-NbSe_2$ through magnetization
measurements.   Each   metastable   vortex    configuration    is
characterized  by  a  critical  current  density  $J_c$  which is
strongly dependent on the magnetic history. It is also shown that
any metastable  vortex  configuration  obtained  under  given
field  historys  can  be  driven  into a stable configuration by
repeated field cycling. This stable configuration has a  critical
current  density  $J_c^{st}$,  uniquely  determined  by  field and
temperature as postulated in a  recent  model\cite{model}.  Field
cycling  appears  to  act  as an effective temperature to drive a
metastable state into the stable state, even when  thermal  energy
itself  is  inadequate  to  sample the phase space and access the
stable state.

The  method  of  recording  minor hysteresis loops described here
allows us to determine the pinning and equilibrium properties  of
the   stable   vortex   state   satisfactorily.  Our  equilibrium
magnetization data clearly suggest that  the  transition  of  the
vortex  lattice  from  an  ordered state to a disordered state is
first order in nature. The smearing of the transition, i.e.,  the
width  of  the  transition  region  may be a manifestation of the
spatially inhomogeneous pinning of  the  system.  The  $J_c^{st}$
data  suggests  that  the  loss  of quasi-long range order in the
vortex  lattice  also  spans  the  same  field  window   as   the
magnetization  jump.  In  the  collective  pinning  picture, this
amounts to correlation volume of the vortex phase  decreasing  in
this  regime  and  the  FLL  becoming completely disordered above
$H_p$ or $T_p$. The precise coincidence of the $J_c$ anomaly with
the equilibrium magnetization  anomaly  further  illustrates  the
self  consistency  of  the  procedure developed here. It would be
interesting  to  compare  the  nature  of  this   disorder-driven
transition  in systems with different types of pinning, e.g. high
density of point pins versus low  density  of  extended  pins  to
further understand the nature of this presumably disorder induced
phase transformation.

\vspace{0.5cm}
The authors thank Dr. K. V. Bhagwat, Dr. T. V. Chandrasekhar Rao,
Dr. P. K. Mishra and Mr. M. R. Singh for discussions. 

\newpage
\begin{table}
\caption{Superconducting parameters in $2H-NbSe_2$}
\begin{tabular}{ccccc}   
T(K)&$H_p$(mT)&$\mu_0\Delta M_{eq}(\mu T)/4\pi$&$J_c(H_p)$(A/m$^2$)
&$\Delta s(k_B)$ \\
\tableline
6.95 & 105  & 3.8$\pm$0.4 & $54 \times 10^4$ &13.6$\pm$1.4  \\
6.90 & 136  & 5.0$\pm$0.4 & $36 \times 10^4$ &13.8$\pm$1.1  \\
6.85 & 170  & 1.3$\pm$0.4 & $26 \times 10^4$ &2.9$\pm$0.9   \\
6.80 & 202  & 1.9$\pm$0.4 & $17 \times 10^4$ &3.5$\pm$0.7   \\
\end{tabular}
\label{table1}
\end{table}
\newpage
\large
\begin{center}
\centerline{{\large {\bf FIGURE CAPTIONS}}}
\end{center}
\normalsize
\noindent
Fig.1.  :  Typical  magnetization hysteresis loop observed in the
peak effect region of  a  superconducting  $2H-NbSe_2$.  In  the
panel (a), the minor curve obtained by decreasing the field from
the point  A, corresponding  to  a  field  ($H  +  \delta$)  on   
the   forward magnetization  curve is shown to saturate at the point
B, which  corresponds to the
field ($H - \delta$). Magnetization values at A and B are $M(H  +
\delta,   \uparrow)$   and   $M_{ML}(H   -  \delta,  \downarrow)$
respectively  (see  text).  In the panel  (b) the  minor  curve 
obtained   by increasing   the   field   from the point  C,  
corresponding  to  a  field
($H-\delta$) on the reverse magnetization curve, saturates
at   D   ($H+\delta$) and  corresponds  to  a  magnetization  value
$M_{ML}(H + \delta,\uparrow)$.

\vspace{0.5cm}
\noindent
Fig.    2:    (a)    Calculated    critical   current   densities
$J_c(H\uparrow)$ and $J_c(H\downarrow)$  in  the  increasing  and
decreasing  field cases, respectively. These are compared with the
stable critical current density $J_c^{st}$ (dotted line). In this
calculation, we have used $H_{low} = 0.05 T$, $H_p = 0.1  T$,  $J_{c1}$
$=$  $10^4  A/m^2$,  $J_{c2}$ $=$ $20 J_{c1}$, $H_1$ $=$ $0.12 T$
and $H_W$  $  =  0.008  T$ [20].  (b)  Magnetization  hysteresis  loop
corresponding  to  the $J_c$ values shown in (a). The hysteresis loop
that would be obtained within the  framework  of  critical  state
model,  i.e., in the limit of $B_r \rightarrow 0$ is also shown in the
in this panel as dotted line. The inset shows the functional form of  $B_r$  
which  is non-zero in the field range $H_{low} < H < H_p$.

\vspace{0.5cm}
\noindent
Fig.  3:  Calculated minor curves of type I and type II are shown
in  pnaels (a) and (b), respectively. In the panel (c), we show the 
$M_{eq}^*$ vs $H$
obtained using Eq. 2  and  Eq.  3,  respectively  along  with  the
original  form,  $M_{eq}  = 0$, assumed in the calculation of the minor
curves.

\vspace{0.5cm}
\noindent
Fig. 4: (a) A part of the magnetization loop (forward and reverse
curves)  measured  at 6.95K on a $2H-NbSe_2$ single crystal. Also
indicated are the characteristic fields, $H_{pl}^+$ and $H_p$.  We
indicate  A  and  B ($H < H_{pl}^+$) and C and D ($H_{pl}^+ < H <
H_p$)  starting  from  which  the  minor  hysteresis  loops   are
measured.  (b)  Minor hysteresis loops started from point A (open
circles). In different field cycles, they are seen to retrace the
same loop. (c) The MHL started from B (continuous  line)  shrinks
with  each  successive field cycle. The increasing and decreasing
field legs of the first and second  cycles  are  numbered.  After
five  field cycles, the hysteresis loop is seen to merge with the
loop shown in (b), which is replotted (open  circles).  (d)  Minor
hysteresis  loops  started  from  point  C (open circles). In the
first field  cycle  itself,  increasing  field  leg  of  the  MHL
moves away   from  the  forward  curve  and  remains  outside  the
magnetization envelope for the subsequent field cycles.  (e)  The
minor  loops  starting  from  D  (continuous  line)  are  seen to
collapse onto the loop shown in (d), which is replotted.

\vspace{0.5cm}
\noindent
Fig.  5:  Magnetization  hysteresis  loop of $2H-NbSe_2$, recorded
using {\it half scan technique} [22] at 6.9K  (continuous  line).  The  open
circles  are  the  saturated  magnetization values obtained after
repeated field cycling. $H_{pl}^+$ and $H_p$ are also marked. The
locus of the saturated magnetization values is shown as a  dotted
line.

\vspace{0.5cm}
\noindent
Fig.  6:  (a)  Stable  critical current density $J_c^{st}$ in the
field range 80 mT $<$ $\mu_0 H$ $<$ 105 mT. In the inset, we  show
the $J_c^{st}$ $vs$ $H$ in the entire field range. Filled
triangles and open circles correspond to the values obtained from
the  MHLs  intiated  from  the  forward and reverse magnetization
curves respectively. (b) Equilibrium magnetization $M_{eq}$ as  a
function  of  field  at  6.95K. Note that the sharp change in $M_{eq}$
coincides with the PE onset field $H_{pl}^+$.

\vspace{0.5cm}
\noindent
Fig.  7:  (a)  Critical current density $J_c^{st}$ vs $H$ and (b)
$M_{eq}$ vs $H$ data obtained at 6.9K. Note that the smeared  jump  in
$M_{eq}$  vs  $H$, as marked by the double sided arrow, agrees precisely
with a smeared jump in critical current density $J_c^{st}$ in the
peak regime. See  text for a discussion.


\begin{references}
\bibitem{note}  For  $H  > H_p$, Eq. 1(b) is more appropriate for
obtaining $M_{eq}$  because  $J_c$  is  independent  of  magnetic
history. This would give $M_{eq} = 0$.
\end{references}
\end{document}